\documentclass[st,twocolumn]{jpsj3}
\usepackage{txfonts}
\usepackage{amsmath}
\usepackage[dvipdfmx]{graphicx}
\graphicspath{{img/}}
\usepackage{color}
\usepackage{siunitx}

\title{Superconducting Quantum Annealing Architecture with LC Resonators}

\author{Hiroto Mukai$^{1,2\dagger}$\thanks{hiroto.mukai@riken.jp, $\dagger$These authors contributed equally to this work.}, Akiyoshi Tomonaga$^{1,2\dagger}$, and Jaw-Shen Tsai$^{1,2}$}
\inst{$^1$Department of Physics, Tokyo University of Science, 1--3 Kagurazaka, Shinjuku, Tokyo 162--0825, Japan\\
$^2$RIKEN, Center for Emergent Matter Science (CEMS), 2--1 Hirosawa, Wako, Saitama 351--0198, Japan\\
} 

\abst{
We propose a novel architecture for superconducting circuits to improve the efficiency of a quantum annealing system. To increase the capability of a circuit, it is desirable for a qubit to be coupled not only with adjacent qubits but also with other qubits located far away. We introduce a circuit that uses a lumped element resonator coupled each with qubit. The resonator-qubit pairs are coupled by rf-superconducting quantum interference device (SQUID)-based couplers. These pairs make a large quantum system for a quantum annealer. This system could prepare a problem Hamiltonian and tune the parameters for quantum annealing.
}

\begin{document}
\maketitle
\section{Introduction}
With the present information explosion in our society, it is indispensable to realize efficient quantum information-processing systems for the coming generation. Such quantum systems are being researched and developed. Among them, superconducting quantum circuits are making remarkable progress. Quantum annealing is a class of quantum information processing specialized for solving optimization problems~\cite{Nishimori1998,Farhi2000,Farhi2001}. In general, a wide range of real-world problems can be classified as optimization problems, which cover the fields of fundamental science, the improvement of productivity, and the development of infrastructure. However, it is practically impossible to solve these optimization problems with von Neumann computers when the size of the problems exceeds certain limits~\cite{Garey1979}.

For quantum annealing, a problem to be solved is encoded as strengths of interactions in a ``spin glass'' that consists of many spins and interactions between spins~\cite{Lucas2014}. By suitably encoding the time evolution of the spin glass, nature itself will find the minimum energy of the whole system, giving us the solution to the optimization problem. To build a quantum annealer, we need to consider what is required for such a physical system. When the number of the interactions for each qubit increases, the encoding of optimization problems becomes more efficient. Therefore, it is possible to reduce the overhead of the number of physical spins~\cite{Boothby2016,Cai2014}. Thus, a larger problem can be solved efficiently. In general, it is indispensable to increase the number of spins as well as couplings between these spins to efficiently solve large-scale problems by quantum annealing. 

In this paper, we propose a novel architecture for scalable quantum annealing circuits with full coupling, in which a spin is coupled to all other spins. The existing superconductive quantum annealing systems~\cite{Harris2010B82,Bunyk2014} utilize flux qubits as spins, which are coupled with each other by an rf-superconducting quantum interference device (SQUID)-based coupler~\cite{Harris2009}. On the other hand, in our proposed architecture, the coupling structure between qubits is mediated by superconducting resonators. Here, the pair of the qubit and resonator functions as a very long quantum system (spin), enabling it to be coupled to a large number of other spins.
A strongly coupled qubit-resonator pair enables us to make a large quantum system compared with the size of a qubit. Additionally, to increase the coupling energy between spins, deep-strong coupling~\cite{Yoshihara2017,Fron2017} between the qubit and the resonator is introduced. It is also possible to introduce a dispersive readout~\cite{Wallraff2005,Jeffrey2014}.

\section{Proposed Architecture}
\begin{figure*}[ht!]
 \centering
 \includegraphics{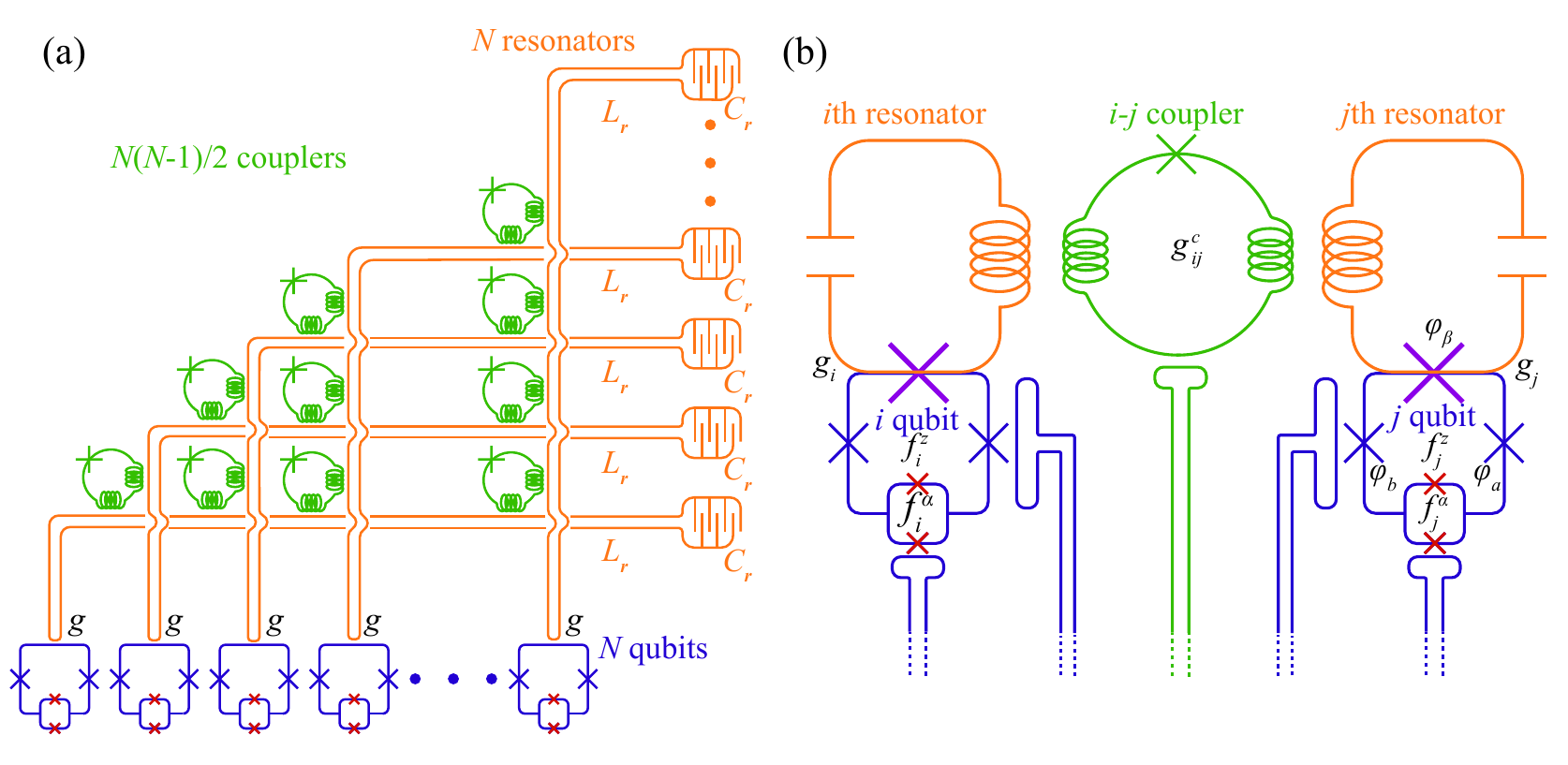}
 \caption{(Color online)
(a)
Schematic representation of full-coupling quantum annealing circuit. The orange part is the LC resonator, which has a long inductive limb.  All qubits (blue part) form a pair with the resonator, and each pair couples with all other pairs via an rf-SQUID-based coupler (green part). $\times$ (crosses) represent Josephson junctions.
(b)
Schematic representation of two flux qubits coupled to two LC resonators via an rf-SQUID-based coupler. The flux qubit and the resonator share a line with a $\beta$-junction. Blue crosses represent Josephson junctions. Red and purple crosses are $\alpha$- and $\beta$-junctions, respectively. $\varphi$ is the phase difference across each junction. $f^\alpha_i$ and $f_i^z$ are flux biases of the $\alpha$-loop and main loop, respectively. Each loop of a qubit and coupler needs control lines.  
For simplicity, these control lines are omitted in the {\it N}-qubit circuit figure (a). In the {\it N}-qubit circuit, control lines and resonators are arranged in a multilayer.
}
 \label{fig:propose}
\end{figure*}

We propose a novel architecture for superconducting circuits to realize a quantum annealing system that consists of flux qubits and lumped element (LC) resonators (Fig.~\ref{fig:propose}).

Each flux qubit has longitudinal (Z) and transverse (X) degrees of freedom~\cite{Mooij1999,Weber2017}. Their Z and X energies are controlled by applied external magnetic fluxes to the main loop and $\alpha$-loop [shown in Fig.~\ref{fig:propose}(b)], respectively. It is common to use flux qubits for quantum annealing because the quadratic structure of the energy band of the flux qubit allows a transverse magnetic field and longitudinal magnetic field to easily and continuously increase or decrease~\cite{Harris2010B81}. For this reason, we also employ flux qubits for our proposed architecture.

In general, a lumped element resonator has a uniform current distribution on its inductive parts, in contrast to a distributed resonator such as those of the coplanar type, with the standing wave dependent on the resonant frequency.
In our architecture, we utilize an LC resonator with a long inductive limb, which plays an important role in our architecture [see Fig.~\ref{fig:propose}(a)]. The long inductive limbs make it possible to couple many spins. Accordingly, an LC resonant mode with a uniform current is realized, while other non-LC resonant modes inevitably exist. However, by optimizing the parameters of the circuit, the energy of the LC resonant mode can be realized in the vicinity of the qubit energy, while making the other modes far away from the energy. Thus, the coupling of the other resonant modes to the qubit can be ignored.

In the architecture, {\it N} flux qubits are arranged on a line and each qubit is connected to a different LC resonator via a mutual inductance. The {\it N} LC resonators are braided so that they fully interact with each other by the long inductive limb through the rf-SQUID-based couplers as shown in Fig.~\ref{fig:propose}(a). Thus, the {\it N} flux qubits effectively and fully interact with all other qubits via the network of LC resonators and couplers.

In contrast with the existing scheme, in which each qubit interacts with some qubits through an rf-SQUID-based coupler, the proposed system has the following advantages. The qubits fully interact with each other. The number of interactions between the qubits is $N(N-1)/2$ when there are $N$ qubits in the proposed circuit. On the other hand, the existing scheme has a unit cell with $2N$ interactions~\cite{Harris2010B82}

In the mapping optimization problems to interactions of a system, the larger the number of spins and interactions between spins, the more efficiently the problems are mapped. Because of the long inductive limbs of resonators for the proposed architecture, it is possible to increase the number of spins and interactions.
%

In the mapping optimization problems to interactions of a system, the larger the number of spins and interactions between spins, the more efficiently the problems are mapped. In our architecture, because of the long inductive limbs of resonators, it is possible to increase the number of spins and interactions.

Considering the actual realization of the quantum annealing circuit, as it is clear from Fig.~\ref{fig:propose}(a), the resonator is interwoven in a stitchlike manner. Therefore, a standard multilayered superconducting integration is required. 
On the other hand, the qubit, which is the part most sensitive to decoherence, can be separately fabricated by the standard double-angle shadow evaporation of aluminum~\cite{Dolan1977} that produces all the good superconducting qubits.

\section{General Quantum Annealing \label{sec:require}}
To perform the quantum annealing, the requirement is that a Hamiltonian of a physical system fits the form of the stoquastic Hamiltonian $(H_{QA})$, which is given by~\cite{Nishimori1998}
\begin{equation}
 \mathcal{H}_{QA} 
= \Lambda(t)\sum_i \tilde{\varepsilon}_i\sigma_i^z 
+ \Lambda(t)\sum_{\substack{i,j \\  i<j }} \tilde{J}_{ij} \sigma^z_i\sigma^z_j
+ \Gamma(t)\sum_i \tilde{\Delta}_i\sigma^x_i ~,
\label{eq:hamiltonianQA}
\end{equation}
where $\tilde{\varepsilon}_i$ and $\tilde{\Delta}_i$ are the normalized energies of the {\it i}th spin corresponding to the longitudinal and transverse magnetic fields, respectively, $\tilde{J}_{ij}$ is the normalized energy of the interaction between the spins $(-1 \le \{ \tilde{\varepsilon}_i,\,\tilde{J}_{ij} \} \le 1)$, and $\Lambda$ and $\Gamma$ are time-dependently tunable values.

The optimization problem is mapped onto $\tilde{\varepsilon}_i$ and $\tilde{J}_{ij}$.
After mapping the optimization problem to the system, the quantum annealing is performed in accordance with the following procedure. Initially $(t=t_0)$, all spins are made to face the same direction by applying transverse magnetic fields, where $\Lambda(t_0)=0$ and $\Gamma(t_0)=1$. Then, the magnetic fields applied to the spins are gradually changed to the longitudinal direction. Finally, at $t=t_f$, $\Lambda(t_f)=1,$ and $\Gamma(t_f)=0$, the states of the spins show us the solution to the problem. When the system is at the end of an annealing procedure, the energy of the spin and the strengths of the effective interactions between spins should be much larger than the transverse energy of the spin $(\tilde{\Delta}_i \ll {\tilde{\varepsilon}_i\,,\tilde{J}_{ij}})$. To satisfy these requirements, the characteristics of our proposed architecture must be estimated and calculated.


\section{Requirements of Proposed Architecture}
The proposed architecture is described by the following Hamiltonian, which considers the qubits, resonators, the longitudinal and transverse inductive coupling  between each qubit and resonator, and the interactions between resonators~\cite{Blais2007}:
\begin{align}
 \mathcal{H}_N = 
 & \sum_{i} 
        \left( \varepsilon_i\sigma_i^z + \Delta_i\sigma_i^x \right)
       +\omega_i^r \left( a_i^\dagger a_i +\frac{1}{2} \right) \notag \\
 & + \sum_{i} 
         g_i^z \sigma^z_i \left(a_i^\dagger +a_i \right)
        +g_i^x \sigma^x_i \left(a_i^\dagger +a_i \right) \notag \\
 & + \sum_{\substack{i,j \\  i<j }}
        g^c_{ij} \left(a_i^\dagger + a_i\right) \left(a_j^\dagger + a_j\right) ~,
\label{eq:hamiltonian}
\end{align}
where {\it i} and {\it j} are integers from 1 to {\it N}, which is the total number of qubits (resonators), 
$\sigma^z_i$ and $\sigma^x_i$ are the {\it i}th spin operators of the longitudinal and transverse degrees of freedom, 
$\varepsilon_i = \varepsilon_i(f_i^z)$ and $\Delta_i=\Delta_i(f_i^\alpha)$ are the energies of each degree of freedom of the {\it i}th qubit, 
$f_i^z =\Phi_i^z/\Phi_0$ and $f_i^\alpha=\Phi^x_i/\Phi_0$, $\Phi_i^z$ and $\Phi_i^x$ are the fluxes of the {\it i}th qubit in the $\alpha$-loop and the main loop, 
$\Phi_0$ is the flux quantum, 
$a_i^\dagger$ and $a_i$ are the bosonic creation and annihilation operators of the {\it i}th resonator, 
$\omega_i^r$ is the energy of the {\it i}th resonator, 
$g_i^z$ and $g_i^x$ are the longitudinal and transverse coupling constants between the {\it i}th qubit and resonator, and $g^c_{ij}$ is the coupling constant between the {\it i}th and {\it j}th resonators, respectively.

When the applied flux of the main loop changes away from a half-integer multiple of the flux quantum, the Z and X energies of the {\it i}th qubit $(\varepsilon_i,\Delta_i)$ become $\Delta_i= 0$ and $\varepsilon_i= \varepsilon_i^f$, and the transverse coupling $g_i^x$ is neglected. The third term of the Hamiltonian $(\mathcal{H}_N)$, which describes the qubit-resonator interactions, is exactly diagonalized (following and expanding Billangeon's method in Refs.~\citen{Billangeon2015} and \citen{Billangeon2015B92}) by the unitary operator given by
\begin{align}
 \mathcal{U}_N = \exp\left[\sum_{k=1}^N \sum_{l=1}^N -\theta_{kl}\sigma_k^z\left(a_l^\dagger - a_l\right)\right] ~.
\end{align}

Here, $\theta_{ij}$ are set to satisfy the following simultaneous equations for all {\it i} and {\it j}:
\begin{equation}
 \forall i,j \quad 
 g_i^z\delta_{ij} - \sum_{k=1}^N \left( 2g_{kj}^c + \omega^r_k\delta_{kj}\right) \theta_{ik} = 0 ~,
 \label{eq:constraint}
\end{equation}
where we impose $g^c_{ij} = g^c_{ji}$ because the coupling strength of the resonators is symmetric, and $g_{ii}=0$ because the self-coupling refers to the self-energy of the resonator $\omega_i^r$, which has already been included.
Under this constraint, using the Baker--Campbell--Hausdorff (BCH) formulation, the first-order terms cancel out, the effective interaction terms remain, and the higher-order terms equal zero,
\begin{align}
 \mathcal{H}'_N 
 \equiv & \mathcal{U}_N^{\dagger}\mathcal{H}_{N}^f\mathcal{U}_{N} \notag \\ 
 = &  \sum_{i}
        \varepsilon_i^f\sigma_i^z
    + \sum_{\substack{i,j \\ i<j}}
        J_{ij} \sigma^z_i \sigma^z_j \notag\\
   &+ \sum_{i}
        \omega_i^r \left( a_i^\dagger a_i +\frac{1}{2} \right)
    + \sum_{\substack{i,j \\ i<j}}
        g^c_{ij} \left(a_i^\dagger + a_i\right) \left(a_j^\dagger + a_j\right)\,, \label{unitaryH} 
\end{align}
where $J_{ij} = -g_i^z \theta_{ji}$.

$J_{ij}$ is the strength of the effective interaction between the {\it i}th and {\it j}th qubits. We can obtain $J_{ij}$ by applying Cramer's rule to Eq.~(\ref{eq:constraint}). $\theta_{ij}$ denotes the route from the {\it i}th qubit to the {\it j}th qubit through the network of resonators. In the proposed architecture, the full interactions between qubits are effectively realized. 
In the Hamiltonian in Eq.~\eqref{unitaryH}, the qubit-resonator interaction term has already been diagonalized.
The remaining terms of resonator couplings change the eigenenergies of the resonators only and not the other terms. 
Therefore, to evaluate the energies of qubits and effective couplings between qubits for the quantum annealing Hamiltonian in Eq.~\eqref{eq:hamiltonianQA}, it is not necessary to consider the 3rd and 4th terms of the Hamiltonian in Eq.~\eqref{unitaryH} that depend only on the resonators themselves.

To map a particular optimization problem that we wish to solve, it is necessary that the set of interactions $(J_{ij})$ is encoded to the set of coupling constants of the system $(g_{ij}^c)$. If the encoding time is not polynomial when using a classical computer, it makes no sense to build the system. In our proposed circuit, once the set of interactions of the problem is fixed, it can be converted to the set of $g^c_{ij}$ by a common matrix method using a classical computer in polynomial time with Eq.~(\ref{eq:constraint}).

We define the coupling matrix $G$ with non-diagonal elements $2g^c_{ij}$ and diagonal elements $\omega^r_i$. Then, using Cramer's rule, we obtain $J_{ij}=-g_i^z (\det G'_{ij}/\det G)$, where $G'_{ij}$ is $G$ with the {\it i}th columns replaced with the vector $(0,\cdots,0,g_j^z,0,\cdots,0)^T$, which has the {\it j}th qubit-resonator coupling strength at the {\it j}th element with other elements equal to zero. The highest orders of $\det G$ and $\det G'_{ij}$ for the resonator energy are 
\begin{align}
 \det G \propto \prod_{k=1}^N \omega^r_k, \qquad \det G'_{ij} \propto 2g^z_jg^c_{ij} \prod_{k=1,\, k\ne i,j}^N \omega_k^r ~.
\end{align}

Therefore, we can estimate the strength of the effective interaction through just two ({\it i} and {\it j}) resonators as
\begin{equation}
 |J_{ij}| \propto (g^z_i/\omega^r_i)(g^z_j/\omega_j^r)g^c_{ij}~,
  \label{eq:strongJij}
\end{equation}
where the lower-order terms of $\omega^r$ are ignored, which are much smaller than the highest-order term because the lower-order terms correspond to coupling through more than two resonators.

In the strong-coupling regime, which is usually used in the field of superconducting circuits, the resonator energy $\omega^r$ is larger than the coupling constant $g$ between a qubit and resonator $(\kappa, \gamma \ll g < \omega^r)$, where $\kappa$ and $\gamma$ are the photon leakage rate from the resonator and the relaxation rate from the qubit, respectively. 
In this regime, the value of $|J_{ij}|$ is much smaller than the sufficient strength of the interactions: $|J_{ij}| \ll 1$.

For example, when $N=2$, $J_{ij}$ is given by
\begin{equation}
 J_{12} = \frac{4g^z_1g^z_2g^c_{12}}{\omega_1^r\omega_2^r - (2g^c_{12})^2} ~.
\label{eq:jij}
\end{equation}
The value of $J_{12}$ is lower than the order of $\mathrm{MHz}$ when common values of the qubit-resonator coupling strength~\cite{Wallraff2004} $(\sim 100\,\mathrm{MHz})$ are used in the strong-coupling regime. To satisfy the requirement of the final procedure (see Sect.~\ref{sec:require}) of the quantum annealing,  $\varepsilon$ must also be lower than the order of $\mathrm{MHz}$. However, the thermal fluctuation of a quantum circuit in a $10\,\mathrm{mK}$ environment is equivalent to a frequency fluctuation of around $200\,\mathrm{MHz}$. Thus, this system cannot give the correct solution to problems because the final state of the qubits is easily buried in thermal noise.

The ultrastrong-coupling regime~\cite{Niemczyk2010,Fron2010} is stronger than the strong coupling regime. However, the coupling strength is still smaller than the energy of the resonator by one order of magnitude $(g \sim 0.1 \omega^r)$, so the strength of the effective interactions is also insufficient in this regime. To resolve this problem of the strength, we adopt the deep-strong-coupling regime, which has recently been realized in experiments~\cite{Yoshihara2017,Fron2017}.

In the deep-strong-coupling regime, the coupling strength between the qubits and resonators is similar to the energy of the resonator $(g \sim \omega^r)$. 
A large inductance, generated by the qubit and the resonator sharing a line and a large Josephson junction (called $\beta$-junction), allows the coupling strength to increase.
In this regime, the strength of the effective interaction is larger and Eq.~(\ref{eq:strongJij}) becomes
\begin{equation}
 |J_{ij}| \propto g^c_{ij} ~.
\end{equation}
The order of the strength of $J_{ij}$ depend the order of the strength of $g_{ij}^c$.

Although in this regime, the approximation of the Jaynes--Cummings model fails, the second-order term of the resonator $g_i^{z,2}(a_i^\dagger + a_i)^2$ appears in the system Hamiltonian without the approximation of the Rabi model. Fortunately, following the method of Ref.~\citen{Yoshihara2017}, this second-order term can be transformed into a first-order term and eliminated to obtain the form of the Hamiltonian in Eq.~(\ref{eq:hamiltonian}).

\begin{figure}[b!]
 \centering
 \includegraphics[width=8cm]{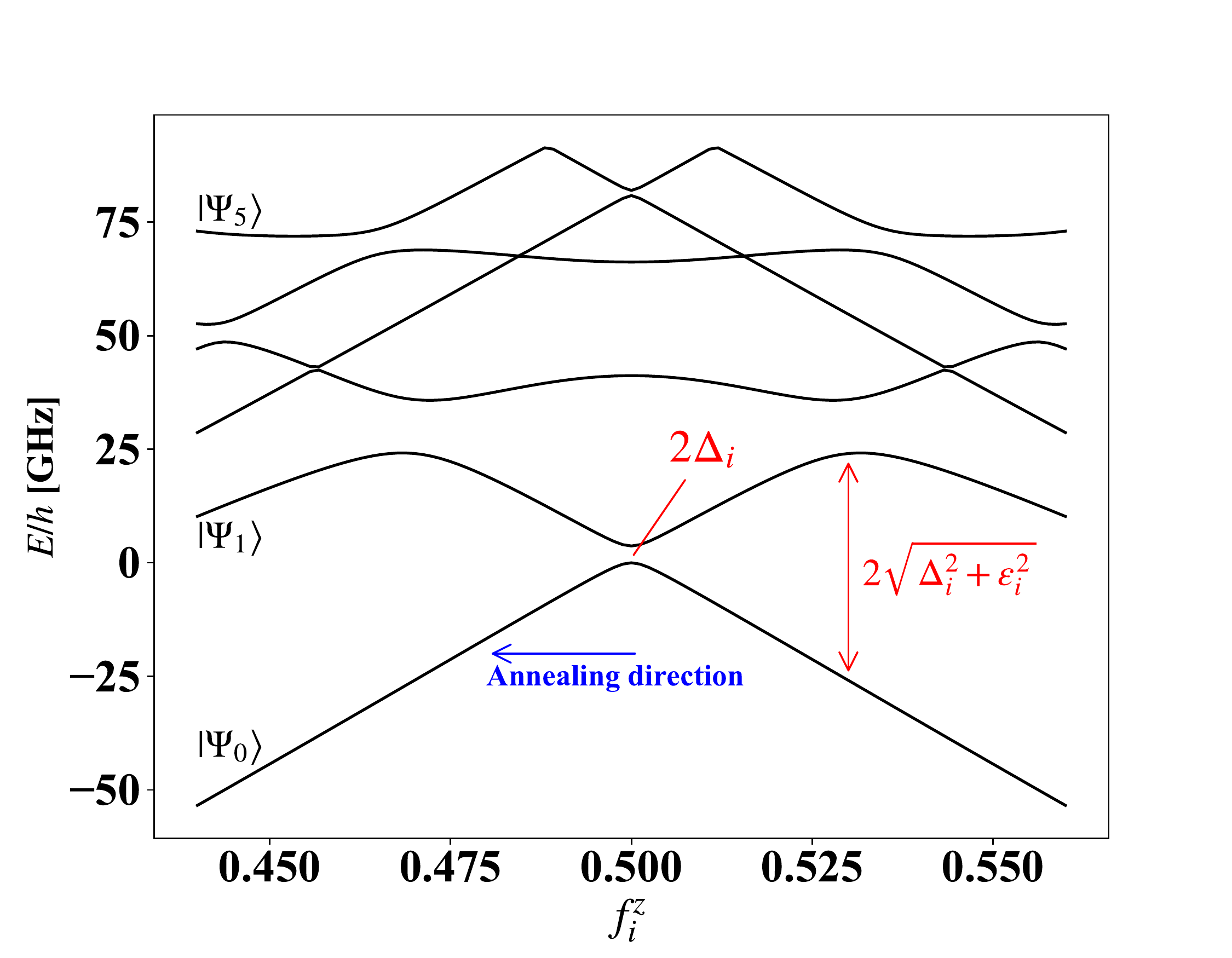}
 \caption{(Color online)
Calculated energy levels of a flux qubit with a $\beta$-junction as a function of main loop flux bias. The energy gap between $E_1$ and $E_0$ is $2\sqrt{\Delta_i^2+\varepsilon_i^2}$. $|\Psi_\xi\rangle$ is the $\xi$th eigenstate given $E_\xi$. We take the calculation space maximum value of $k,l,$ and $m$ of 7 for a good approximation. The parameters are $E_c/h=5\,\mathrm{GHz}$, $E_J/h=250\,\mathrm{GHz}$, $\alpha=0.7$, and $\beta=4$.
}
 \label{fig:energy}
\end{figure}

Next, we calculate the energy levels and the coupling strengths of the qubit in the deep-strong-coupling regime.
From Fig.~\ref{fig:propose}(b), the Hamiltonian of the {\it i}th qubit is given by
\begin{align}
 \mathcal{H}^q_i = 
 & \frac{1}{2(\alpha + \beta + 2\alpha\beta)C}
        \left[(\alpha + \beta + \alpha\beta)\left(q_a^2+q_b^2 \right) \right. \notag \\ 
 & \quad\quad \left. -(1+2\alpha)q_\beta^2-2\alpha \beta q_a q_b-2\alpha(q_a+q_b)q_\beta \right]\notag \\
 & - E_J \left\{ \cos\varphi_a +\cos\varphi_b+\cos\varphi_\beta \right. \label{eq:hamiltonianQ} \\
 & \quad\quad \left. +\alpha\cos\left(\pi f^\alpha_i \right) \cos\left[ \varphi_a+\varphi_b+\varphi_\beta+\pi(2f_i^z -f^\alpha_i) \right] \right\} ~, \notag
\end{align}
where $q_j$ is the conjugate momentum of $\phi_j=\varphi_j \Phi_0/2\pi,\,j\in \{a,\,b,\,\beta\}$, $C$ is the capacitance of the Josephson junction, $E_J$ is the energy of the Josephson junction, and $0.5\alpha E_J$ and $\beta E_J$ are the energies of the $\alpha$-junction and $\beta$-junction, respectively. To derive the energy levels, we calculate the Schr\"odinger equation of the Hamiltonian $(\mathcal{H}^q_i)$ by using the wave function $|\Psi_\xi\rangle =\sum_{k,l,m} C_{k,l,m}^\xi | \psi_k^{(a)}\rangle | \psi_l^{(b)}\rangle | \psi_m^{(\beta)} \rangle$, where $| \psi_\eta^{(j)}\rangle = (2\pi)^{-1/2}\exp(-i\eta\varphi_j)$, $C^\xi_{k,l,m}$ is an arbitrary complex number for $\eta\in \{k,l,m\}$ and $\xi$ is the number of energy states. The energy band structure is shown in Fig.~\ref{fig:energy}.

\begin{figure}[b!]
 \centering
 \includegraphics[width=8cm]{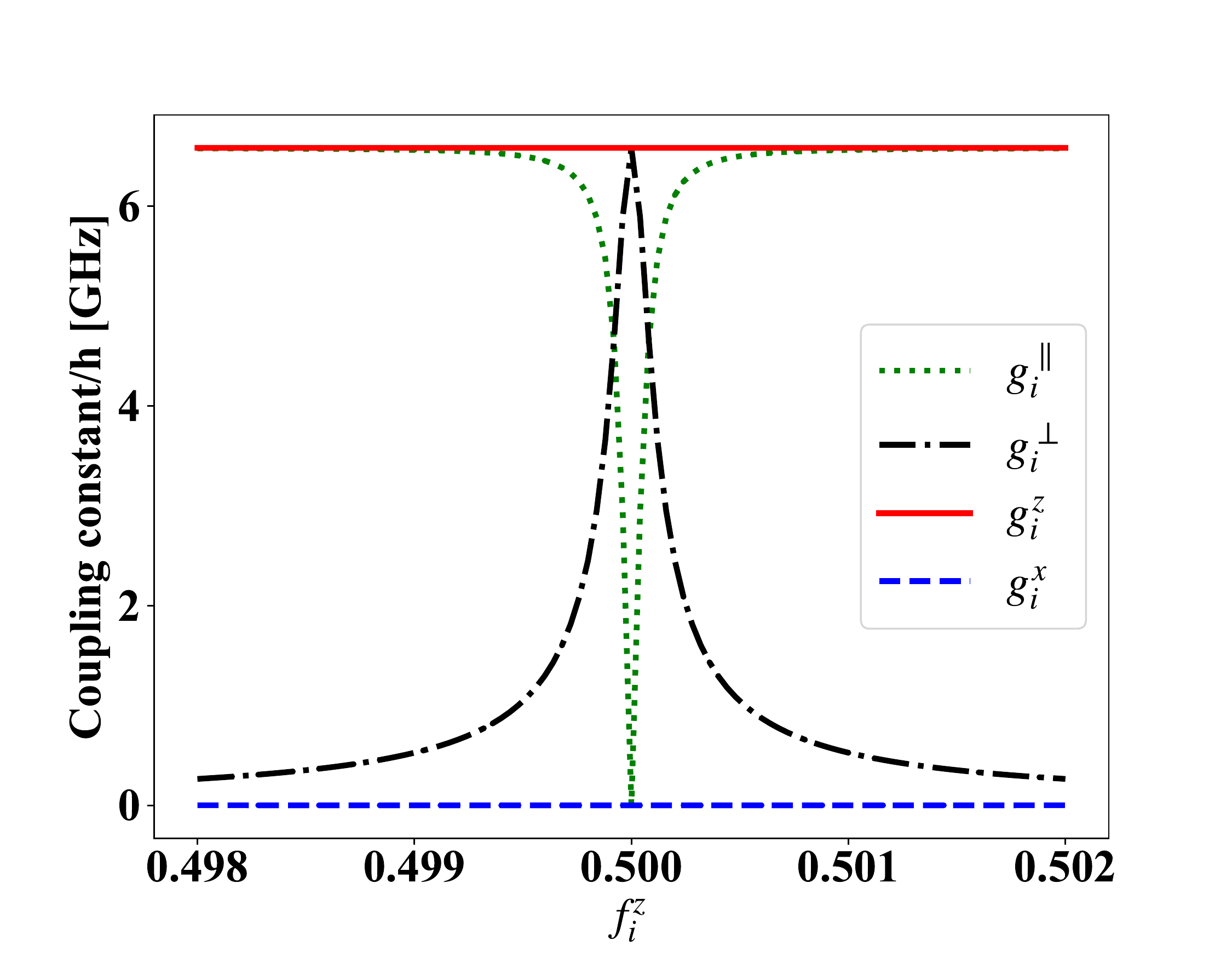}  
 \caption{(Color online)
Calculated coupling constant of each degree of freedom as a function of the main loop flux bias. The solid and dashed curves represent $g_i^z$ and $g_i^x$, respectively. These constants are calculated from $g_i^\parallel$, $g_i^\perp$, $\varepsilon_i$, and $\Delta_i$. The other parameters are shown in Table~\ref{table:para}.
}
 \label{fig:coupling}
\end{figure}

The coupling constant between the qubit and the resonator via the $\beta$-junction is also calculated~\cite{Peropadre2013} (shown in Fig. \ref{fig:coupling}) as
\begin{equation}
 g_i^\parallel = \frac{1}{2}I_r \times \frac{1}{2}\Phi_0 \left( \left\langle \Psi_1\right|\varphi_\beta\left|\Psi_1\right\rangle - \left\langle \Psi_0\right|\varphi_\beta\left|\Psi_0 \right\rangle \right) ~,
\label{eq:zcoupling}
\end{equation}
\begin{equation}
 g_i^\perp = \frac{1}{2}I_r \times \frac{1}{2}\Phi_0 \left( \left\langle \Psi_0\right|\varphi_\beta\left| \Psi_1\right\rangle + \left\langle \Psi_1\right|\varphi_\beta\left| \Psi_0 \right\rangle \right) ~,
\label{eq:xcoupling}
\end{equation}
where $\varphi_\beta$ is the phase difference at the $\beta$-junction.

As shown in Fig.~\ref{fig:energy}, a flux qubit can be well approximated as a two-level system around the optimal point $(f_i^z \sim 0.5)$ because of its large anharmonicity~\cite{Orland1999}. Using the Hamiltonian in Eq.~\eqref{eq:hamiltonianQ} and both of the coupling constants in Eqs.~\eqref{eq:zcoupling} and \eqref{eq:xcoupling}, the Hamiltonian of the resonator-qubit pair is given by 
\begin{align} 
 \hspace{-2.4mm}\mathcal{H}_i^{qr} \hspace{-1mm}
 =\omega_i^r \left(a_i^\dagger a_i + \frac{1}{2} \right) + \omega_i^q \sigma_i^{z'} \hspace{-1mm}+ \left(g_i^\parallel \sigma^{z'}_i + g_i^\perp \sigma_i^{x'} \right)\left( a_i^\dagger +a_i \right), 
\label{eq:hamiltonianpair} 
\end{align} 
where $\omega_i^q=\sqrt{\Delta_i^2+\varepsilon_i^2}$ and the Pauli matrix $\sigma_i^{z'}$ basis is $|\Psi_0\rangle$ and $|\Psi_1\rangle$. $g_i^z$ and $g_i^x$ in the coordinate $\sigma_i^z$ of the  Hamiltonian in Eq.~\eqref{eq:hamiltonian} can be calculated using $g_i^\parallel$, $g_i^\perp$, $\varepsilon_i$, and $\Delta_i$ (Fig.~\ref{fig:coupling}). Thereby, $g^x_i$ is negligible because it is much smaller than $g_i^z$.

In the deep-strong-coupling regime, the state of the pair of the qubit and resonator is displaced. When the qubit energy is sufficiently smaller than the resonator energy, the state of the pair at the transverse magnetic field can be approximated~\cite{Nori2010} to $\left| \leftarrow \right\rangle_i \equiv \left| + \right\rangle_i \otimes \exp\left[-(g_i^z/\omega^r_i)(a_i^\dagger -a_i)\right] \left| n_i \right\rangle$. Here, $\left| + \right\rangle_i \equiv (\left| \circlearrowleft \right\rangle_i + \left| \circlearrowright \right\rangle_i)/\sqrt{2}$, $\left| \circlearrowleft \right\rangle_i$ and $\left| \circlearrowright \right\rangle_i$ are the basis of the {\it i}th qubit and correspond to the current directions, and $n_i$ is the photon number of the Fock state in the {\it i}th resonator.

\begin{figure}[t!]
 \centering
 \includegraphics[width=8cm]{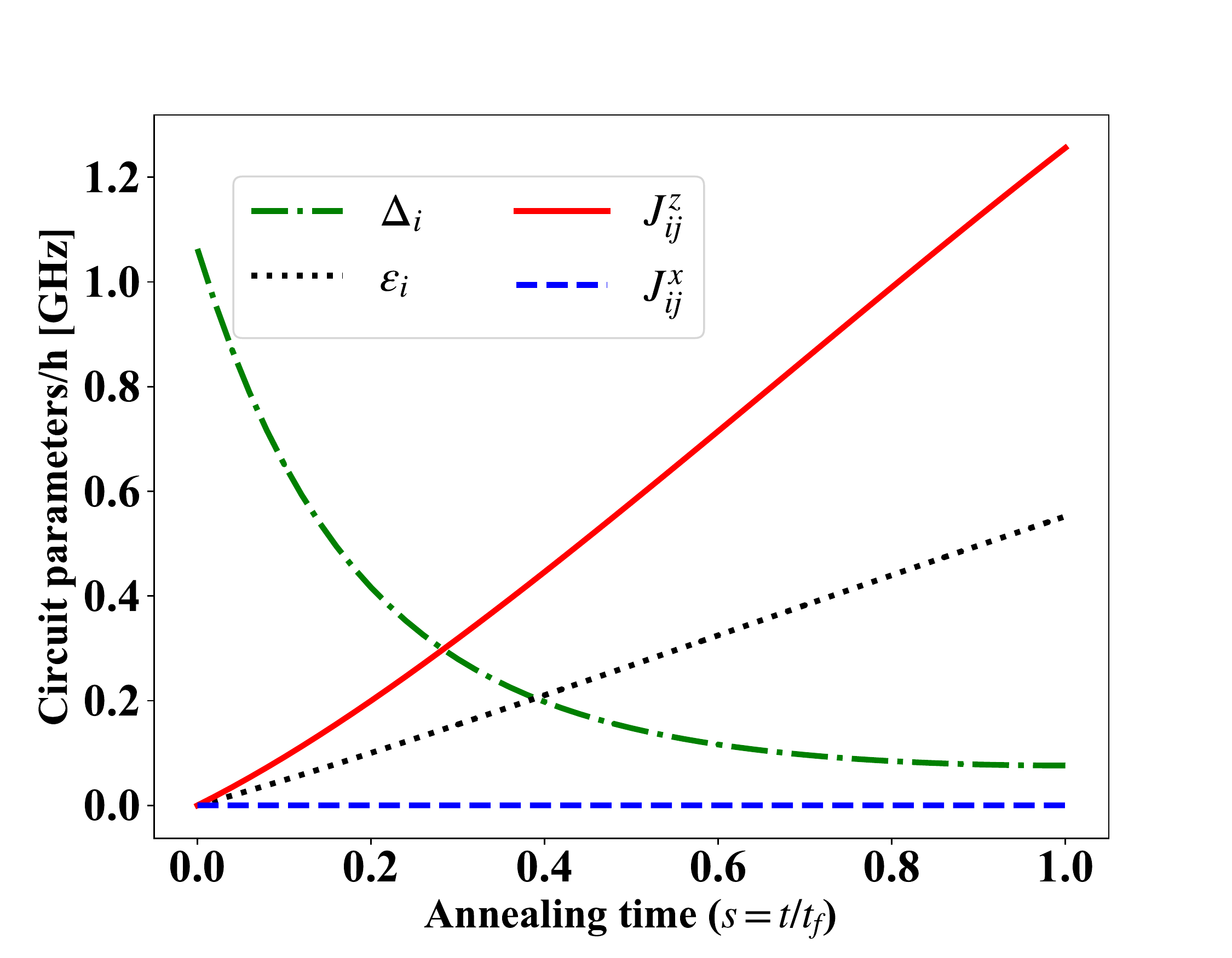}  
 \caption{(Color online)
Calculated circuit parameter dynamics for a typical annealing path. This graph is based on the assumption that $g^c_{ij}$ is linearly increased from 0 to 400 MHz. $f_i^z$ and $f^\alpha_i$ are also linearly changed from 0.5 to 0.4997 and from 0.21 to 0, respectively. During the annealing path, $J_{ij}^z$ (solid curve) and $J_{ij}^x$ (dashed curve) are calculated from $g_i^z$ and $g_i^x$ in Eq.~(\ref{eq:hamiltonian}). The other parameters are shown in Table~\ref{table:para}.
}
 \label{fig:ap}
\end{figure}

Next, we describe a procedure to perform quantum annealing using the parameters in the proposed circuit $(\Delta_i,\,\varepsilon_i,\,g_i^z,\,g_i^x,\,g^c_{ij})$. An example of the procedure for each parameter during annealing is shown in Fig.~\ref{fig:ap}. This graph is based on the assumption that the flux biases are linearly changed at each loop.

In the beginning of the quantum annealing procedure, the state of each qubit is $\left| \leftarrow \right\rangle$. Then, to fit the Hamiltonian of the proposed architecture [Eq.~\eqref{eq:hamiltonian}] to the form of Eq.~\eqref{eq:hamiltonianQA}, $J_{ij},\,\Delta_i,$ and $\varepsilon_i$ must be controlled with time. To control the parameters, they are time-dependently tuned by external flux biases. The four parameters $(\Delta_i,\,\varepsilon_i,\, J_{ij}^z,\,J_{ij}^x)$ depend on the flux biases of the main loop and the $\alpha$-loop because $J_{ij}$ depends on $g_{i,j}^z$, $g_{i,j}^x$, and the set of $g_{ij}^c$. Such a standard annealing path is shown in Fig.~\ref{fig:ap}.

In the proposed system, we can freely choose $\varepsilon_i$ and $|J_{ij}^z|$ in the range of 0 to around $2\,\mathrm{GHz}$. At the end of the path, $\Delta_i$ should be set much smaller than $\varepsilon_i$ and $|J_{ij}^z|$~\cite{Grajcar2005}. Fortunately, $\Delta_i$ is reduced by the factor $\exp[-2(g_i^z/\omega_i^r)^2]$ in the deep-strong-coupling regime~\cite{Nori2010}. Therefore, the final state of the system should correspond to the solution to an optimization problem. After the annealing, the flux qubit can be measured by dispersive readout with high accuracy. 

From Eq.~\eqref{eq:jij}, the coupling strength ($J_{12}$) is expressed in terms of the circuit parameters as
\begin{align}
 J_{12}\approx \frac{M^2}{L_r^2} M_c I^q_1I^q_2~,
\label{eq:j12}
\end{align}
where $M$ is the mutual inductance between a qubit and a resonator, and $M_c$ is the effective mutual inductance between resonators (1 and 2) through the coupler. $I^q_i$ is the screening current of the {\it i}th qubit ($i=1,2$). $L_r$ is the effective inductance of resonators. 

To realize antiferromagnetic and ferromagnetic interactions between qubits for the mapping of problems in Eq.~\eqref{eq:jij}, the rf-SQUID-based coupler connecting the resonators requires that the coupling must be able to take positive and negative values under external biases~\cite{Harris2009,Wulschner2016}.
To meet the requirement from the coupler, the circuit parameters are chosen to obtain the coupling strength ($J_{12}$) on the order of $\mathrm{GHz}$. The parameters are listed in Table~\ref{table:para}.

Parasitic direct couplings exist between resonators because of geometric mutual inductance at their limbs. However, in the case that two resonators with the parameters given in Table~\ref{table:para} are positioned 100\,\si{\micro\meter} apart, the parasitic direct couplings should be lower than the order of $\mathrm{MHz}$. The simulation showed that such small parasitic couplings can be ignored. When the length of the resonator limb is elongated to the order of $\mathrm{cm}$, the parasitic couplings probably need to be suppressed with a superconducting ground plane.

An {\it N}-qubit system can clearly be realized in the same way. 
In this {\it N}-pair circuit, we can show that the order of coupling strengths is not reduced by the increase in {\it N}.
We deal with this {\it N}-qubit system as a unit cell because the number of qubits in the unit cell is limited by the length of the long inductive limb of the LC resonators. From Eq.~(\ref{eq:j12}), to make $J_{ij}$ as large as possible, the inductance of the resonator cannot be made too large. It is necessary to suppress $L_r$ to $\mathrm{nH}$ order or below to construct a fully coupled circuit with dozens of qubits. For this reason, the length of the long inductive limb is limited to $\mathrm{cm}$ order. 

\begin{table}[t!]
  \caption{
  Parameters for annealing calculation. $I^r_i$ is the root-mean-square current of the resonator. $I_c$ and $I_s$ are critical currents of Josephson junctions in an rf-SQUID-based coupler and a qubit, respectively. $E_J=I_c\Phi_0/2\pi, E_c=e^2/2C$. The resonator with the parameters below has a 2-\si{\milli\metre}-long inductor limb.}
  \label{table:data_type}
  \centering
  \begin{tabular}{ccc}
    \hline
    Parameter   & Value & Unit  \\
    \hline \hline
    $E_c/h$     & 5     & GHz   \\
    $E_J/h$     & 250   & GHz   \\
    $\omega_r$  & 7.2   & GHz   \\
    $I_r$       & 41    & nA    \\
    $L_r$       & 1.4   & nH    \\
    $M_c$       & 154   & pH    \\
    $I_s$       & 10    & \si{\micro\meter} \\
    $\alpha$    & 0.8   & -     \\
    $\beta$     & 1.1   & -     \\
    \hline
     \label{table:para}
  \end{tabular}
\end{table}

\section{Conclusion}
We have described the architecture of a quantum annealing circuit with lumped element resonators to considerably increase the number of coupled qubits, which is important for an efficient quantum annealing system. Although the total number of fully coupled qubits is considered to be limited to around 100 in a unit at present, this unit can be scaled up by combining with other units via other couplers. 

Quantum annealing machines with dozens of fully coupled qubit unit cells should have an obvious advantage in mapping problems such as social networks, economics, and categorized advertisements. These problems can be decomposed into many subsets, where tight relationships exist within the subset, while only shallow relationships are required among subsets. 

\section*{Acknowledgment}
We thank  Z. H. Peng, Y. Zhou, R. Wang, D. Zhang, F. Yoshihara, T. Yamamoto, M. Maezawa, S. Watabe, T. Nikuni, K. Imafuku, S. Kawabata, Y. Matsuzaki, Y. Seki, E. Otani, Y. Nakajima, and K. Sakata for their thoughtful comments on this research. 
This paper is based on results obtained from a project commissioned by the New Energy and Industrial Technology Development Organization (NEDO). Support from CREST, JST (Grant No. JPMJCR1676), and the ImPACT Program of Council for Science, Technology and Innovation (Cabinet Office, Government of Japan) is also appreciated.

\end{document}